\documentclass[a4paper]{jpconf}
\usepackage[utf8]{inputenc}
\usepackage{graphicx}
\begin{document}

  \newcommand {\nc} {\newcommand}
  \nc {\beq} {\begin{eqnarray}}
  \nc {\eeq} {\nonumber \end{eqnarray}}
  \nc {\eeqn}[1] {\label {#1} \end{eqnarray}}
  \nc {\mrm} [1] {\mathrm{#1}}
  \nc {\bce}{\begin{center}}
  \nc {\ece} {\end{center}}
  \nc {\ex} [1] {\ensuremath{^{#1}}}
  \nc {\btrs} {\begin{tabular*}}
  \nc {\etrs} {\end{tabular*}}
  \nc {\btr} {\begin{tabular}}
  \nc {\etr} {\end{tabular}}
  \nc {\half} {\mbox{$\frac{1}{2}$}}
  \nc {\thal} {\mbox{$\frac{3}{2}$}}
  \nc {\fial} {\mbox{$\frac{5}{2}$}}
  \nc {\ve} [1] {\mbox{\boldmath $#1$}}
  \nc {\eq} [1] {(\ref{#1})}
  \nc {\Eq} [1] {Eq.~(\ref{#1})}
  \nc {\Ref} [1] {Ref.~\cite{#1}}
  \nc {\Refc} [2] {Refs.~\cite[#1]{#2}}
  \nc {\Sec} [1] {Sec.~\ref{#1}}
  \nc {\chap} [1] {Chapter~\ref{#1}}
  \nc {\anx} [1] {Appendix~\ref{#1}}
  \nc {\tbl} [1] {Table~\ref{#1}}
  \nc {\fig} [1] {Fig.~\ref{#1}}
  \nc {\Sch} {Schr\"odinger }
  \nc {\flim} [2] {\mathop{\longrightarrow}\limits_{{#1}\rightarrow{#2}}}
  \nc {\textdegr}{$^{\circ}$}

\title{Mechanisms of direct reactions with halo nuclei}

\author{\underline{P.~Capel}\ex{1}, F.~M.~Nunes\ex{2}, H.~Esbensen\ex{3},
R.~C.~Johnson\ex{4}}

\address{\ex{1}Physique Nucléaire et Physique Quantique (CP 229), Université Libre de Bruxelles (ULB),\\
50 av. F. D. Roosevelt, B-1050 Brussels, Belgium\\
\ex{2}National Superconducting Cyclotron Laboratory
and Department of Physics and Astronomy, Michigan State University,
East Lansing, MI 48824, USA\\
\ex{3}Physics Division, Argonne National Laboratory, Argonne, IL 60436, USA\\
\ex{4}Department of Physics, University of Surrey, Guildford GU2 7XH, UK}

\ead{pierre.capel@ulb.ac.be}




\begin{abstract}
Halo nuclei are exotic nuclei which exhibit a strongly clusterised structure:
they can be seen as one or two valence nucleons loosely bound to a core.
Being observed at the ridge of the valley of stability, halo nuclei are studied mostly through reactions.
In this contribution the reaction models most commonly used
to analyse experimental data are reviewed and compared to one another.
A reaction observable built on the ratio of two angular distributions is then presented.
This ratio enables removing most of the sensitivity to the reaction mechanism,
which emphasises the effects of nuclear structure on the reaction.
\end{abstract}

\section{Introduction}
The advent of radioactive-ion beams has enabled the exploration
of the nuclear landscape far from stability.
This technological breakthrough has led to the discovery
of exotic nuclear structures, such as halo nuclei \cite{Tan85l}.
Halo nuclei exhibit a large matter radius in comparison with their isobars.
This peculiarity is qualitatively explained by the small binding energy of one or two valence nucleons,
which then have a significant probability of presence at a large distance from the core of the nucleus.
They thus exhibit a strongly clusterised structure in which
one or two loosely-bound nucleons form a diffuse halo around a dense core.
Examples of one-neutron halo nuclei are \ex{11}Be and \ex{15}C,
whereas \ex{6}He and \ex{11}Li are the best known two-neutron halo nuclei.
\ex{8}B is a candidate one-proton halo nucleus.

Being far away from stability, halo nuclei cannot be studied through usual
spectroscopic techniques, and we have to rely on indirect methods,
such as reactions, to infer information about their structure.
Breakup is probably the mostly used reaction to study
halo nuclei \cite{Fuk04,Nak09}. In that reaction, the halo dissociates from the core through the interaction
with a target. In order to get valuable information from experimental data, various reaction
models have been developed: the Continuum Discretised Coupled Channel model (CDCC) \cite{TNT01},
the Time-Dependent approach (TD) \cite{EBB95}, and the Dynamical Eikonal Approximation (DEA) \cite{BCG05}.
These models are presented in \Sec{model}. That section also includes a comparison between
them for the breakup of \ex{15}C on Pb at $68A$MeV \cite{CEN12},
which has been measured at RIKEN \cite{Nak09}.

To reduce the sensitivity of the measurements to the reaction mechanism,
one can also look for an observable that is independent of the reaction mechanism.
Such an observable is presented in \Sec{ratio}. It corresponds to the ratio of two angular distributions.
According to the prediction of the Recoil Excitation and Breakup model (REB) \cite{JAT97},
this ratio should depend only on the projectile structure.
This prediction is confirmed by DEA calculations of reactions involving \ex{11}Be,
the archetypical one-neutron halo nucleus \cite{CJN11}.

\section{Reaction models}\label{model}
\subsection{Theoretical framework}
Let us consider the reaction involving a one-neutron halo nucleus,
in which a valence neutron $f$ is loosely bound to a core $c$.
Such a two-cluster projectile is described by the Hamiltonian
\beq
H_0=T_r+V_{cf}(\ve{r}),
\eeqn{e1}
where $\ve{r}$ is the neutron-core relative coordinate,
$T_r$ is the $c$-$f$ kinetic-energy operator,
and $V_{cf}$ is a real potential that simulates the interaction between the halo neutron
and the core.
The parameters of this phenomenological potential are adjusted to reproduce
the binding energy of the projectile and some of its low-lying levels.
The structure of the target $T$ is usually neglected and its interaction with the projectile components
is simulated by optical potentials $V_{cT}$ and $V_{fT}$. Within this framework, the theoretical
study of reactions involving halo nuclei reduces to solving the three-body \Sch equation
\beq
\left[T_R+H_0+V_{cT}(R_{cT})+V_{fT}(R_{fT})\right]\Psi(\ve{r},\ve{R})=E_T\Psi(\ve{r},\ve{R}),
\eeqn{e2}
where $\ve{R}$ is the projectile-target relative coordinate and
$T_R$ is the corresponding kinetic-energy operator.
\Eq{e2} is solved with the initial condition that the projectile is in its ground state $\phi_0$
\beq
\Psi(\ve{r},\ve{R})\flim{Z}{-\infty}e^{iKZ+\ldots}\phi_0(\ve{r}),
\eeqn{e3}
where $K$ is the wave number of the initial relative motion of the projectile to the target,
which is assumed to be in direction $\ve{\widehat{Z}}$.
The reaction models differ mostly in the way \Eq{e2} is solved.

In the CDCC model, \Eq{e2} is solved by expanding $\Psi$ upon the eigenstates of $H_0$ \cite{TNT01}.
Since breakup leads to the dissociation of the projectile, a tractable description
of the projectile continuum must be included in this expansion.
That is why the continuum is \emph{discretised}.
This is usually achieved by dividing the continuum
into small energy intervals called \emph{bins}.
Each bin wave function is obtained by averaging the wave functions of
the continuum states within the corresponding energy interval.
This leads to a set of coupled equations, which we solve using {\sc fresco} \cite{fresco}.
Apart from the continuum discretisation, the CDCC technique solves \Eq{e2} exactly. It is
a purely quantal model with no approximation on the projectile-target relative motion,
and it does not imply any restriction on the beam energy.
However, it is very expensive in a computational point of view,
in particular at high beam energy.

A simpler way to solve \Eq{e2} is the Time-Dependent (TD) approach,
which relies on a semiclassical approximation.
In that approximation, the projectile-target
relative motion is simulated by a classical trajectory $\ve{R}(t)$.
Along that trajectory, the projectile is affected by the time-varying Coulomb
and nuclear fields of the target. The evolution of its internal structure is
then obtained by solving a \emph{time-dependent \Sch equation} \cite{EBB95}.
That equation must be solved for each possible trajectory,
with the condition that the projectile is initially in its bound state.
Various computer codes have been developed to numerically solve the time-dependent equation.
In this study, we will use the one detailed in \Ref{EBB95}.
Thanks to its semi-classical approximation, the time-dependent method is faster than CDCC.
However, due to this approximation, quantal interferences between the trajectories cannot
be taken into account.

Another way to avoid the heavy CDCC machinery is to use the eikonal approximation.
At sufficiently high energy, one can assume that the projectile-target relative motion will be
close to the initial plane wave \eq{e3}. The main idea of the eikonal approximation is to factorise that
plane wave out of the three-body wave function $\Psi$ and to assume that the remaining factor
is smoothly varying with $\ve{R}$. This leads to the Dynamical Eikonal Approximation (DEA),
whose equation is mathematically equivalent to a time-dependent \Sch equation for straight-line trajectories \cite{BCG05}.
As in the TD approach, the DEA equation must be solved for all the possible ``trajectories''.
However, it remains a quantal approximation since it does not rely on any semiclassical approximation.
This implies in particular that, contrarily to the TD approach,
the DEA accounts for the quantal interferences between different ``trajectories''.
However, since it relies on the eikonal approximation,
the DEA is valid only at intermediate and high energies.

\subsection{{\rm \ex{15}C} Coulomb breakup}
\begin{figure}[h]
\begin{minipage}{19pc}
\center
\includegraphics[width=18pc]{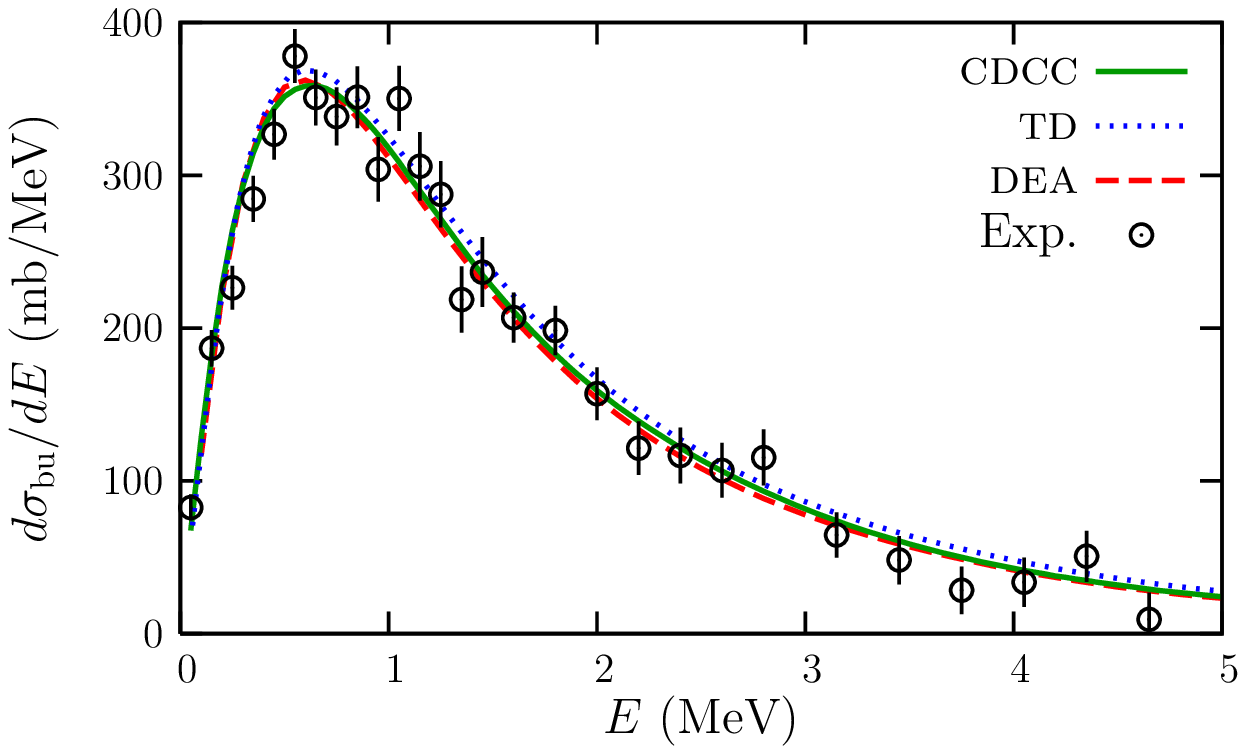}
\caption{Energy distribution for the breakup of \ex{15}C on Pb at $68A$MeV.
Comparison of CDCC, TD, and DEA \cite{CEN12}. Data from \Ref{Nak09}}\label{f1}
\end{minipage}\hspace{1pc}%
\begin{minipage}{18pc}
\center
\includegraphics[width=17.5pc]{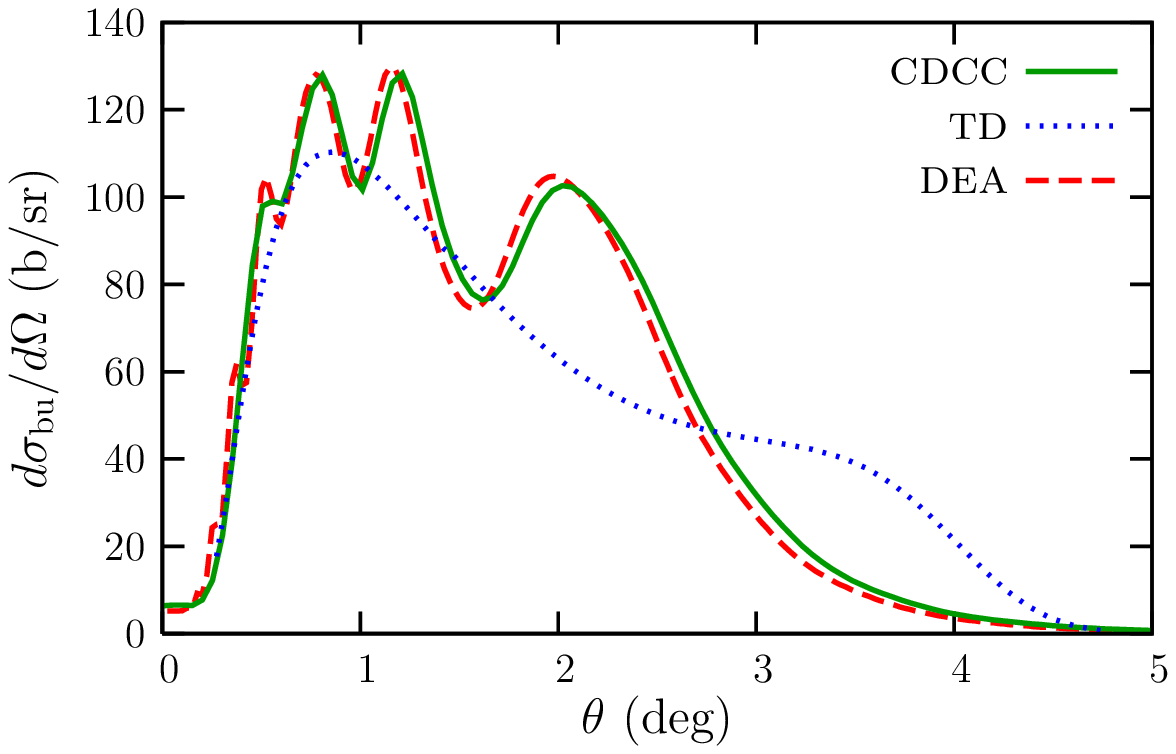}
\caption{Angular distribution for the breakup of \ex{15}C on Pb at $68A$MeV.
Comparison of CDCC, TD, and DEA \cite{CEN12}.}\label{f2}
\end{minipage} 
\end{figure}

To compare the three aforementioned models, we consider the breakup of the one-neutron halo nucleus
\ex{15}C on lead at $68A$MeV \cite{Nak09}.
The breakup cross section is plotted in \fig{f1} as a function of the relative energy $E$
between the halo neutron and the \ex{14}C core after dissociation.
All three models are in excellent agreement with each other \cite{CEN12}
and with experiment \cite{Nak09}.
This result shows that the reaction process is well understood, and that \ex{15}C exhibits
a two-cluster structure.

\fig{f2} shows the breakup angular distribution, i.e. the breakup cross section as a function
of the scattering angle $\theta$ of the \ex{14}C-n centre of mass \cite{CEN12}.
CDCC and DEA lead to very similar cross sections. In particular they exhibit the same interference
pattern, which confirms the validity of the DEA at these intermediate energies, even though
it is much less time consuming than CDCC.
As expected, the TD technique does not exhibit any quantal interference. Nevertheless,
it reproduces the general trend of the quantal angular distributions,
which explains why, once integrated over $\theta$, it gives reliable results.

\section{Ratio technique}\label{ratio}
Albeit reliable, the aforementioned models remain sensitive to their input parameters.
In particular, the choice
the optical potentials $V_{cT}$ and $V_{fT}$ can affect significantly the results of
the calculations \cite{CGB04}.
An observable less sensitive to the reaction mechanism would naturally eliminate
such sensitivity and improve the analysis of exotic nuclear structures from reactions.

It has been observed that angular distributions for elastic scattering and breakup exhibit
very similar features: Coulomb rainbow, oscillatory patterns etc.\ \cite{CHB10}.
This result can be qualitatively understood within the Recoil
Excitation and Breakup model (REB) \cite{JAT97}.
That model relies on two simplifying assumptions.
First it includes an adiabatic treatment of the excitation of the projectile, i.e.
the projectile excitation energy is neglected compared to the beam energy.
Second, it neglects the interaction between the halo neutron and the target $V_{fT}$.
Under these two assumptions an elegant factorisation for the elastic-scattering
cross section is obtained \cite{JAT97}
\beq
\frac{d\sigma_{\rm el}}{d\Omega}=|F_{0,0}|^2\left(\frac{d\sigma}{d\Omega}\right)_{\rm pt},
\eeqn{e14}
which is the product of a cross section for the elastic scattering
for a pointlike projectile $\left(d\sigma/d\Omega\right)_{\rm pt}$
times a form factor, which accounts for the halo extension
\beq
F_{0,0}=\int |\phi_0(\ve{r})|^2 e^{i\ve{Q}\cdot\ve{r}}d\ve{r},
\eeqn{e15}
where $\ve{Q}$ is proportional to the transferred momentum.
A similar factorisation can be performed for the angular distribution
for the breakup of the projectile to energy $E$ in the continuum
\beq
\frac{d\sigma_{\rm bu}}{dEd\Omega}=|F_{E,0}|^2\left(\frac{d\sigma}{d\Omega}\right)_{\rm pt},
\eeqn{e16}
where the form factor reads
\beq
|F_{E,0}|^2=\sum_{ljm}\left|\int \phi_{ljm}(E,\ve{r})\phi_0(\ve{r}) e^{i\ve{Q}\cdot\ve{r}}d\ve{r}\right|^2.
\eeqn{e17}
It contains, besides the wave function of the initial bound state
of the projectile $\phi_0$, the wave functions describing the
broken up projectile in partial wave $ljm$ at energy $E$ in the continuum.

\begin{figure}
\bce
\includegraphics[width=25pc]{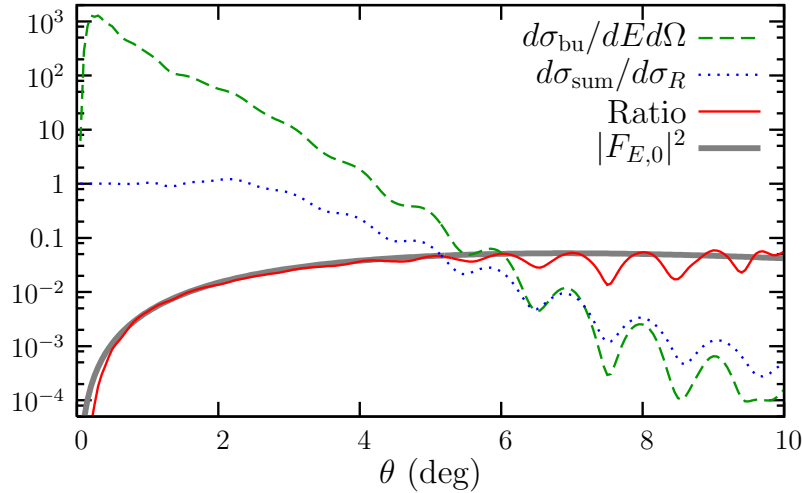}
\ece
\caption{DEA calculations of \ex{11}Be impinging on Pb at $69A$MeV.
Angular distribution for breakup at $E=0.1$~MeV (dashed line) and
the summed cross section (dotted line) are shown with their ratio (solid line) and
the REB prediction (thick gray line) \cite{CJN11}.}\label{f3}
\end{figure}

The factorisations \eq{e14} and \eq{e16} explain the results of \Ref {CHB10} since both expressions
contain the same $\left(d\sigma/d\Omega\right)_{\rm pt}$, which is responsible for
most of their angular dependence.
In addition, the REB model also provides the idea for the \emph{ratio technique}:
within that model, the ratio of cross sections \eq{e14} and \eq{e16} is the ratio
of form factors \eq{e15} and \eq{e17}, which depend only on the projectile structure.
Accordingly, the ratio should be independent of the reaction process.
In order to test this idea, we perform reaction calculations within the DEA \cite{BCG05},
which does not include an adiabatic treatment of the projectile excitation,
and which includes $V_{fT}$.
As a testcase, we consider the collision of \ex{11}Be with Pb at $69A$MeV,
which correspond to the conditions of a RIKEN experiment \cite{Fuk04}
with which the DEA calculations are in excellent agreement \cite{GBC06}.
The result of this test is summarised in \fig{f3}, which presents the DEA angular distributions,
their ratio, and its REB prediction \cite{CJN11}.
Note that instead of the mere elastic-scattering cross section \eq{e14}, we use
the summed cross section $d\sigma_{\rm sum}/d\Omega$,
which corresponds to all quasi-elastic processes
(elastic and inelastic scattering and elastic breakup).
It is plotted here as a ratio to the Rutherford cross section (dotted line).
This gives the best agreement between DEA and REB \cite{CJN11}.
As already noted in \Ref{CHB10}, both angular distributions exhibit similar features.
Interestingly their ratio (solid line) does not exhibit such features and it behaves
smoothly with the scattering angle $\theta$. In addition the DEA ratio is in excellent agreement
with its REB prediction (thick grey line).
This result suggests the ratio idea to be an excellent reaction observable to study loosely-bound
systems, such as halo nuclei.
Further tests have confirmed its independence to the reaction mechanism:
the ratio computed for a carbon target is nearly superimposed to the one obtained on lead \cite{CJN11}.
Moreover, the ratio is strongly sensitive to the binding energy, orbital, and radial wave function
of the valence neutron \cite{CJN11}, indicating that valuable information about the projectile
structure can be inferred from this observable.

\section{Summary and prospect}
Halo nuclei are among the most peculiar nuclear structures discovered thanks to the development
of radioactive-ion beams. They exhibit a strongly clusterised structure, which explains their
large matter radius in comparison to their isobars.
Being located on the shoulder of the valley of stability, they are studied mostly through
reactions, such as elastic scattering and breakup.
Therefore, accurate reaction models must
be available to extract valuable information from reaction data.
In this contribution the mostly used reaction models have been presented
and compared for the breakup of \ex{15}C, a typical halo nucleus, at intermediate energy:
the Continuum Discretised Coupled Channel model (CDCC) \cite{TNT01},
the Time-Dependent approach (TD) \cite{EBB95},
and the Dynamical Eikonal Approximation (DEA) \cite{BCG05}.
Although based on different assumptions, they lead to nearly identical
cross sections, but for the quantal interferences, which cannot be reproduced within TD,
due to its relying on the semiclassical approximation.

To remove the influence of the reaction mechanism on observables,
we suggest to measure the ratio of angular distributions, e.g. for scattering
and breakup. As predicted by the REB model \cite{JAT97}, this new observable
is nearly independent of the reaction mechanism and provides detailed information
on the halo structure. It seems that the idea could be easily applied to Borromean
neutral systems like \ex{11}Li.
Extensions to charged valence nucleon, such as proton hal\oe s
may also be possible.
We believe this ratio method will open a new era in the study of exotic systems.


\section*{References}
\begin{thebibliography}{9}
\bibitem{Tan85l} I.~Tanihata \etal,  Phys. Rev. Lett. \textbf{55}, 2676 (1985).

\bibitem{Fuk04} N.~Fukuda \etal, Phys. Rev. C \textbf{70}, 054606 (2004).

\bibitem{Nak09} T.~Nakamura \etal, Phys. Rev. C \textbf{79}, 035805 (2009).

\bibitem{TNT01} J.~A.~Tostevin, F.~M.~Nunes, and I.~J.~Thompson, Phys. Rev.  C \textbf{63}, 024617 (2001).

\bibitem{EBB95} H.~Esbensen, G.~F.~Bertsch, and C.~A.~Bertulani, Nucl. Phys. \textbf{A581}, 107 (1995).

\bibitem{BCG05} D.~Baye, P.~Capel, and G.~Goldstein, Phys. Rev. Lett. \textbf{95}, 082502 (2005).

\bibitem{CEN12} P.~Capel, H.~Esbensen, and F.~M.~Nunes, Phys. Rev. C \textbf{85}, 044604 (2012).

\bibitem{JAT97} R.~C.~Johnson, J.~S.~Al-Khalili, and J.~A.~Tostevin,
Phys. Rev. Lett. \textbf{79}, 2771 (1997).

\bibitem{CJN11} P.~Capel, R.~C.~Johnson, and F.~M.~Nunes , Phys. Lett. \textbf{B705}, 112 (2012).

\bibitem{fresco} I. J. Thompson, Comput. Phys. Rep. \textbf{7}, 167 (1988).

\bibitem{CGB04} P.~Capel, G.~Goldstein, and D.~Baye, Phys. Rev. C \textbf{70}, 064605 (2004).

\bibitem{CHB10} P.~Capel, M.~Hussein, and D.~Baye, Phys. Lett. \textbf{B693}, 448 (2010).

\bibitem{GBC06} G.~Goldstein, D.~Baye, and P.~Capel, Phys. Rev. C \textbf{73}, 024602 (2006).

\end{thebibliography}
\end{document}